\documentclass[]{aa}
\usepackage[varg]{txfonts}

\usepackage{natbib}
\bibpunct{(}{)}{;}{a}{}{,} 
\usepackage{graphicx}
\usepackage{color}
\usepackage[colorlinks,citecolor=blue]{hyperref}
\usepackage{amstext}
\usepackage{lscape}

\begin{document}

\title{The electron energy-loss rate due to radiative recombination}

\author{Junjie Mao\inst{\ref{inst1},~\ref{inst2}}, 
        Jelle Kaastra\inst{\ref{inst1},~\ref{inst2}}
        \and N. R. Badnell\inst{\ref{inst3}}}

\offprints{J. Mao,~\email{J.Mao@sron.nl}}

\institute{SRON Netherlands Institute for Space Research, Sorbonnelaan 2, 
           3584 CA Utrecht, the Netherlands \label{inst1} 
           \and Leiden Observatory, Leiden University, Niels Bohrweg 2, 
           2300 RA Leiden, the Netherlands \label{inst2}
           \and Department of Physics, University of Strathclyde, 
           Glasgow G4 0NG, UK \label{inst3}}

\date{Received date / Accepted date}

\abstract
{For photoionized plasmas, electron energy-loss rates 
due to radiative recombination (RR) are required for thermal equilibrium calculations, 
which assume a local balance between the energy gain and loss.
While many calculations of total and/or partial RR rates are available from literature, 
specific calculations of associated RR electron energy-loss rates are lacking.
}
{Here we focus on electron energy-loss rates
due to radiative recombination of H-like to Ne-like ions 
for all the elements up to and including zinc ($Z=30$), 
over a wide temperature range.}
{We use the AUTOSTRUCTURE code to calculate 
the level-resolved photoionization cross section
and modify the ADASRR code 
so that we can simultaneously obtain level-resolved RR rate coefficients and
associated RR electron energy-loss rate coefficients.
The total RR rates and electron energy-loss rates of 
\ion{H}{i} and \ion{He}{i} are compared with those found in literature.
Furthermore, we utilize and parameterize 
the weighted electron energy-loss factors (dimensionless) 
to characterize total electron energy-loss rates due to RR.}
{The RR electron energy-loss data are archived according to 
the Atomic Data and Analysis Structure (ADAS) data class {\it adf48}.
The RR electron energy-loss data are also incorporated 
into the SPEX code for detailed modelling of photoionized plamsas. }
{}
\keywords{atomic data -- atomic processes}

\titlerunning{RR electron energy-loss rate}
\authorrunning{Mao et al.}
\maketitle

\section{Introduction}
\label{sct:intro}
Astrophysical plasmas observed in the X-ray band can roughly be divided into 
two subclasses: collisional ionized plasmas and photoionized plasmas. 
Typical collisional ionized plasmas include stellar coronae 
(in coronal/collisional ionization equilibrium), 
supernova remnants (SNRs, in non-equilibrium ionization)
and the intracluster medium (ICM). 
In a low-density, high-temperature collisional ionized plasma, e.g. ICM, 
collisional processes play an important role \citep[for a review, see e.g.][]{kaa08}.
In contrast, in a photoionized plasma, photoionization, recombination 
and fluorescence processes are also important in addition to collisional processes. 
Both the equations for the ionization balance
(also required for a collisional ionized plasma)
and the equations of the thermal equilibrium are used to 
determine the temperature of the photoionized plasma. 
Typical photoionized plasmas in the X-ray band
can be found in X-ray binaries (XRBs) and active galactic nuclei (AGN). 

For collisional ionized plasmas, various calculations of 
total radiative cooling rates are available in the literature, 
such as \citet{cox71}, \citet{ray76}, \citet{sut93}, 
\citet{sch09}, \citet{fos12} and \citet{lyk13}. 
These calculations take advantage of full plasma codes like 
SPEX \citep{kaa96} and APEC \citep{smi01}, 
and do not treat individual energy-loss (cooling) processes separately. 
Total radiative cooling rates include 
the energy-loss of both the line emission and the continuum emission.
The latter includes the energy-loss due to radiative recombination (RR).
Even more specifically, the energy-loss due to RR can be 
separated into the electron energy-loss and the ion energy-loss. 

On the other hand, for photoionized plasmas, 
the electron energy-loss rate due to RR is 
one of the fundamental parameters for thermal equilibrium calculations, 
which assume a local balance between the energy gain and loss. 
Energy can be gained via photoionization, Auger effect, 
Compton scattering, collisional ionization, collisional de-excitation and so forth. 
Energy-loss can be due to radiative recombination, dielectronic recombination, 
three body recombination, inverse Compton scattering, collisional excitation,
bremsstrahlung, etc., as well as the line/continuum emission following these atomic processes.
In fact, the energy-loss/gain of all these individual processes need to be known.
The calculations of electron energy-loss rates due to RR in the Cloudy code 
\citep{fer98,fer13} are based on hydrogenic results \citep{fer92,lam01}.
In this manuscript, we focus on improved calculations of 
the electron energy-loss due to radiative recombination,
especially providing results for He-like to Ne-like isoelectronic sequences.

While several calculations of RR rates, 
including the total rates and/or detailed rate coefficients, 
for different isoelectronic sequences are available, 
e.g. \citet{gu03} and \citet{bad06},
specific calculations of the associated electron energy-loss rate 
due to RR are limited. The pioneering work was done by \citet{sea59}
for hydrogenic ions using the asymptotic expansion of the Gaunt factor 
for photoionization cross sections (PICSs).

By using a modified semi-classical Kramers formula 
for radiative recombination cross sections (RRCSs), 
\citet{kim83} calculated the total RR electron energy-loss rate 
for a few ions in a relatively narrow temperature range.

\citet{fer92} used the $nl$-resolved hydrogenic PICSs provided by 
\citet{sto91} to calculate both $n$-resolved RR rates ($\alpha_{\rm i}^{\rm RR}$) 
and electron energy-loss rates ($L_{\rm i}^{\rm RR}$). 
Contributions up to and include $n=1000$ are taken into account.

Using the same $nl$-resolved hydrogenic PICSs provided by \citet{sto91},
\citet{hum94} calculated the RR electron energy-loss rates for hydrogenic ions 
in a wide temperature range. In addition, \citet{hum98} calculated 
PICSs of \ion{He}{i} (photoionizing ion) for $n \leq 25$ with their close-coupling 
\textbf{\textit{R}}-matrix calculations.
Together with hydrogenic \citep{sto91} PICSs for $n > 25$
(up to $n=800$ for low temperatures), 
the RR electronic energy-loss rate coefficient of \ion{He}{i} 
(recombined ion) was obtained.

Later, \citet{lam01} used the exact PICSs from the Opacity Project \citep{sea92} 
for $n < 30$ and PICSs of \citet{ver96} for $n \ge 30$ to obtain
$n$-resolved RR electron energy-loss rates for hydrogenic ions 
in a wide temperature range.
The authors introduced the ratio of $\beta / \alpha$ (dimensionless), 
with $\beta = L / kT$ and $L$ the RR electron energy-loss rate.
The authors also pointed out that  $\beta / \alpha$ changes merely by 1 dex 
in a wide temperature range meanwhile $\alpha$ and $\beta$ change more than 12 dex.

In the past two decades, more detailed and accurate calculations of PICSs
of many isoelectronic sequences have been carried out \citep[e.g.][]{bad06},
which can be used to calculate specifically the electron energy-loss rates due to RR.

Currently, in the SPEX code \citep{kaa96},
the assumption that the mean kinetic energy of a recombining electron is 
$3kT/4$ \citep{kal82} is applied for calculating the electron energy-loss rate due to RR.
Based on the level-resolved PICSs provided by the 
AUTOSTRUCTURE\footnote{http://amdpp.phys.strath.ac.uk/autos/} 
code \citep[v24.24.3,][]{bad86}, the electron energy-loss rates due to RR 
are calculated for the H-like to Ne-like isoelectronic sequences 
for elements up to and including Zn ($Z=30$) in a wide temperature range.
Subsequently, the electron energy-loss rate coefficients ($\beta = L / kT$)
are weighted with respect to the total RR rates ($\alpha_{\rm t}$), 
yielding the weighted electron energy-loss factors ($f= \beta / \alpha_{\rm t}$, 
dimensionless). The weighted electron energy-loss factors can be used, 
together with the total RR rates, to update the description
of the electron energy-loss due to RR in the SPEX code or other codes.

In Sect.~\ref{sct:mo}, we describe the details of the numerical calculation 
from PICSs to the electron energy-loss rate due to RR. 
Typical results are shown graphically in Sect.~\ref{sct:res}. 
Parameterization of the weighted electron energy-loss factors 
is also illustrated in Sect.~\ref{sct:par}.
The detailed RR electron energy-loss data are archived according to 
the Atomic Data and Analysis Structure (ADAS) data class {\it adf48}.
Full tabulated (unparameterized and parameterized) 
weighted electron energy-loss factors are available in CDS.
Comparison of the results for \ion{H}{i} and \ion{He}{i} can be found 
in Section~\ref{sct:cf_0101_0202}. 
The scaling of the weighted electron energy-loss factors 
with respect to the square of the ionic charge of the recombined ion 
can be found in Section~\ref{sct:sca_z2}. 
We also discuss the electron and ion energy-loss due to RR 
(Section~\ref{sct:rrc}) and the total RR rates (Section~\ref{sct:rr_tot}).

Throughout this paper, we refer to the recombined ion
when we speak of the radiative recombination of a certain ion, 
since the line emission following the radiative recombination 
comes from the recombined ion.
Furthermore, only RR from the ground level of the recombining ion 
is discussed here. 

\section{Methods}
\label{sct:mo}
\subsection{Cross sections}
\label{sct:cs}
The AUTOSTRUCTURE code is used for calculating
 level-resolved non-resonant PICSs 
under the intermediate coupling (``IC") scheme \citep{bad03}. 
The atomic and numerical details can be found in \citet{bad06}, 
we briefly state the main points here.
We use the Slater-Type-Orbital model potential to determine the radial functions.
PICSs are calculated first at zero kinetic energy of the escaping electron, 
and subsequently on a $z$-scaled logarithmic energy grid with three points per decade,
ranging from $\sim z^2 10^{-6}$ to $z^2 10^2$ ryd, where $z$
is the ionic charge of the photoionizing ion/atom. 
PICSs at even higher energies are at least several orders of magnitude smaller 
compared to PICSs at zero kinetic energy of the escaping electron. 
Nonetheless, it still can be important, especially for the $s$- and $p$-orbit, 
to derive the RR data at the high temperature end. 
We take advantage of the analytical hydrogenic PICSs 
\citep[calculated via the dipole radial integral,][]{bur65}
and scale them to the PICS with the highest energy calculated by 
AUTOSTRUCTURE to obtain PICSs at very high energies.
Note that fully $nLSJ$-resolved PICSs for those levels 
with $n\le15$ and $l\le3$ are calculated specifically. 
For the rest of the levels, we use the fast, 
accurate and recurrence hydrogenic approximation \citep{bur65}.
Meanwhile, bundled-$n$ PICSs for 
$n=$~16,~20,~25,~35,~45,~55,~70,~100,
~140,~200,~300,~450,~700,~999  are also calculated specifically
in order to derive the total RR and electron energy-loss rates
(interpolation and quadrature required as well).

The inverse process of dielectronic and radiative recombination is
resonant and non-resonant photoionization, respectively. 
Therefore, radiative recombination cross sections (RRCSs) 
are obtained through the Milne relation under the principle of detailed balance 
(or microscopic reversibility) from non-resonant PICSs. 

\subsection{Rate coefficients}
\label{sct:rc}
The RR rate coefficient is obtained by 
\begin{equation}
\small
\alpha_{i}(T) = \int_0^{\infty} v~\sigma_{i}(v)~f(v,~T)~dv~,
\label{eq:rr_def}
\end{equation}
where $v$ is the velocity of the recombining electron, $\sigma_{i}$
is the individual detailed (level/term/shell-resolved) RRCS,
$f(v,~T)$ is the probability density distribution of 
the velocity of the recombining electrons
for the electron temperature $T$. 
The Maxwell-Boltzmann distribution for the free electrons 
is adopted throughout the calculation, 
with the same quadrature approach as described in \citet{bad06}. 
Accordingly, the total RR rate per ion/atom is 
\begin{equation}
\small
\alpha_{\rm t}(T) =  \sum_{i} \alpha_{i}(T)~.
\label{eq:rr_ioa}
\end{equation}
Total RR rates for all the isoelectronic sequences,
taking contributions up to $n=10^3$ into account 
(see its necessity in Section~\ref{sct:res}).

The RR electron energy-loss rate coefficient is defined as \citep[e.g.][]{ost89}
\begin{equation}
\small
\beta_{i}(T) = \frac{1}{kT}~\int_0^{\infty} \frac{1}{2}~m~v^3~\sigma_{i}(v)~f(v,~T)~dv~,
\label{eq:el_def}
\end{equation}
The total electron energy-loss rate due to RR is obtained simply by adding  
all the contributions from individual captures, 
\begin{equation}
\small
L_{\rm t}(T) = \sum_{i}~L_i = kT~\sum_{i}~\beta_i~,
\end{equation}
which can be identically derived via
\begin{equation}
\small
L_{\rm t}(T) = kT~\alpha_{\rm t}(T)~f_{\rm t}(T)~,
\label{eq:el_ioa}
\end{equation}
where 
\begin{equation}
\small
f_{\rm t}(T) = \frac{\sum_{i}~\beta_{i}(T)}{\alpha_{\rm t}(T)}~,
\label{eq:beta_ioa}
\end{equation}
is defined as the weighted electron energy-loss factor (dimensionless) hereafter. 

The above calculation of the electron energy-loss rates is realized by 
adding Equation~(\ref{eq:el_def}) into the archival post-processor FORTRAN code 
ADASRR\footnote{http://amdpp.phys.strath.ac.uk/autos/ver/misc/adasrr.f} (v1.11).
Both the level-resolved and bundled-$n/nl$ RR data
and the RR electron energy-loss data are obtained.
The output files have the same format of {\it adf48}
with RR rates and electron energy-loss rates in the units of 
${\rm cm^3~s^{-1}}$ and ${\rm ryd~cm^3~s^{-1}}$, respectively.
Note that ionization potentials of the ground level of the recombined ions 
from NIST\footnote{http://physics.nist.gov/PhysRefData/ASD/ionEnergy.html} 
(v5.3) are adopted to correct the conversion from PICSs to RRCSs 
at low kinetic energy for low-charge ions.
We should point out that although the level-resolved 
and bundled-$nl$/$n$ RR data are, in fact, available on 
OPEN ADAS\footnote{http://open.adas.ac.uk/adf48}, 
given the fact that we use the latest version of the AUTOSTRUCTURE code
and a modified version of the ADASRR code, 
here we re-calculate the RR data, which are used together with the 
RR electron energy-loss data to derive the weighted electron energy-loss factor 
$f_{\rm t}$ for consistency. 
In general, our re-calculate RR data are almost identical to those on OPEN ADAS, 
except for a few many-electron ions at the the high temperature end, 
where our re-calculated data differ by a few percent. 
Whereas, both RR data and electron energy-loss data are 
a few orders of magnitude smaller compared to those at the lower temperature end,
thus, the above mentioned difference has negligible impact on
the accuracy of the weighted electron energy-loss factor
(see also in Section~\ref{sct:rr_tot}). 

For all the isoelectronic sequences discussed here, the conventional ADAS
19-point temperature grid $z^2(10-10^7)$~K is used.

\section{Results}
\label{sct:res}
\begin{figure}
\centering
\includegraphics[width=\hsize]{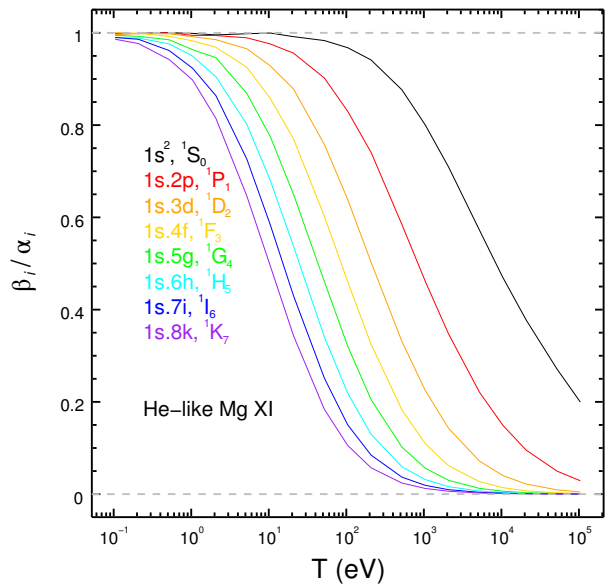}
\caption{For He-like \ion{Mg}{xi}, the ratio between 
level-resolved electron energy-loss rates $L_i$ 
and the corresponding radiative recombination rates 
times the temperature of the plasma, 
i.e. $\beta_i / \alpha_i$ (not be confused with $\beta_i / \alpha_{\rm t}$), where
$i$ refers to the $nLSJ$-resolved levels with $n \le 8$ 
(shown selectively in the plot). }
\label{fig:rr_welf_0212}
\end{figure}

For each individual capture due to radiative recombination, 
when $kT \ll I$, where $I$ is the ionization potential, 
the RR electron energy-loss rate $L_i$ is nearly identical to $kT~\alpha_i$,
since the Maxwellian distribution drops exponentially for $E_{\rm k} \gtrsim kT$, 
where $E_{\rm k}$ is the kinetic energy of the free electron before recombination.
On the other hand, when $kT \gg I$, 
the RR electron energy-loss rate is negligible compared with $kT~\alpha_i$.
As in an electron-ion collision, when the total energy in the incident channel 
nearly equals that of a closed-channel discrete state, the channel interaction 
may cause the incident electron to be captured in this state \citep{fan68}.
That is to say, those electrons with $E_{\rm k} \simeq I$ are preferred to be captured,
thus, $L_i \sim I~\alpha_i$.  Figure~\ref{fig:rr_welf_0212} shows the ratio of
$\beta_i / \alpha_i = L_i / (kT \alpha_i )$ for representative 
$nLSJ$-resolved levels (with $n \le 8$) of He-like \ion{Mg}{xi} . 

\begin{figure}
\centering
\includegraphics[width=\hsize]{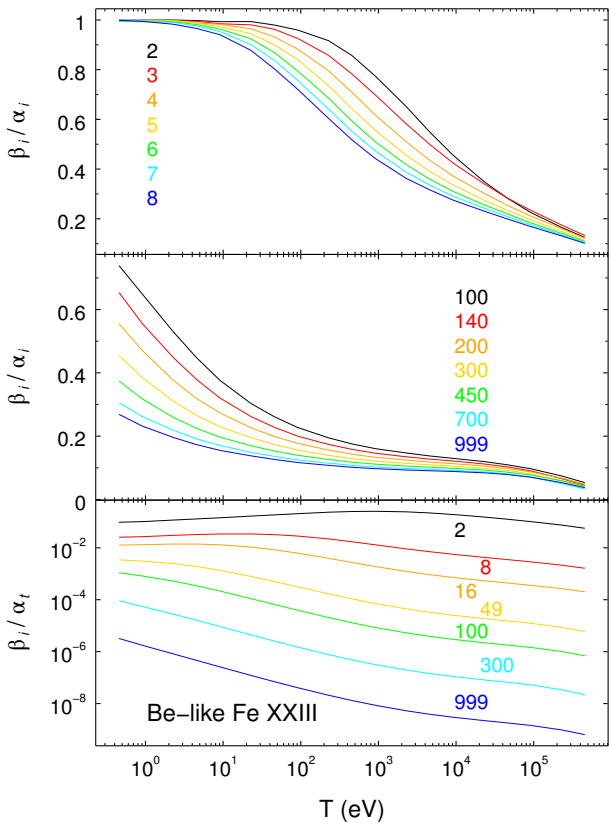}
\caption{Ratios of $\beta_i / \alpha_i$ for Be-like \ion{Fe}{xxiii} 
(upper and middle panel) and ratios of $\beta_i / \alpha_{\rm t}$ 
(bottom panel) where $i$ refers to the shell number. 
Low- and high-$n$ shell results are shown selectively in the plot.
The upper panel shows all the shells with $n \le 8$.
The middle panel shows shells with 
$n=$100,~140,~200,~300,~450,~700 and 999. 
In the lower panel the shells are $n=2,~8,~16,~49,~100,~300,~999$.
}
\label{fig:rr_welf_0426}
\end{figure}

In terms of capturing free electrons into individual shells (bundled-$n$),  
due to the rapid decline of the ionization potentials for those very high-$n$ shells,
the ionization potentials can be comparable to $kT$, 
if not significantly less than $kT$, at the low temperature end. 
Therefore we see the significant difference between 
the top panel (low-$n$ shells) and middle panel (high-$n$ shells) 
of Figure~\ref{fig:rr_welf_0426}.
In order to achieve adequate accuracy, 
contributions from high-$n$ shells (up to $n \le 10^3$) ought to be included.
The middle panel of Figure~\ref{fig:rr_welf_0426} shows clearly that 
even for $n=999$ (the line at the bottom), at the low temperature end, 
the ratio between $\beta_{n=999}$ and $\alpha_{n=999}$ 
does not drop to zero. Nevertheless, the bottom panel of 
Figure~\ref{fig:rr_welf_0426} illustrates the advantage of 
weighting the electron energy-loss rate coefficients with respect to 
the total RR rates, i.e. $\beta_{i} / \alpha_{\rm t}$, 
which approaches zero more quickly.
At least, for the next few hundreds shells following $n=999$, 
their weighted electron energy-loss factors should be no more than
$10^{-5}$, thus, their contribution to the total electron energy-loss rate 
should be less than 1\%. 

\begin{figure}
\centering
\includegraphics[width=\hsize]{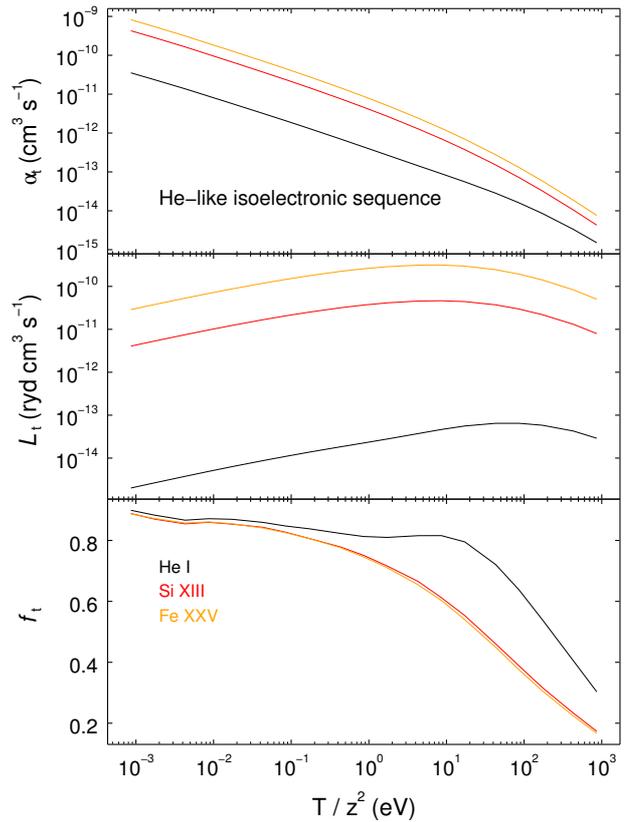}
\caption{The total RR rates $\alpha_{\rm t}$ (top),
electron energy-loss rates $L_{\rm t}$ (middle)
and weighted electron energy-loss factors $f_{\rm t}$ (bottom)
of He-like isoelectronic sequences for ions including
\ion{He}{i} (black), \ion{Si}{xiii} (red) and \ion{Fe}{xxv} (orange).
The temperature is down-scaled by $z^2$, where $z$ is the ionic charge 
of the recombined ion, to highlight the discrepancy between hydrogenic and non-hydrogenic. 
The captures to form the \ion{He}{i} shows non-hydrogenic feature in the bottom panel.
}
\label{fig:rr_welf_02ies}
\end{figure}

\begin{figure}
\centering
\includegraphics[width=\hsize]{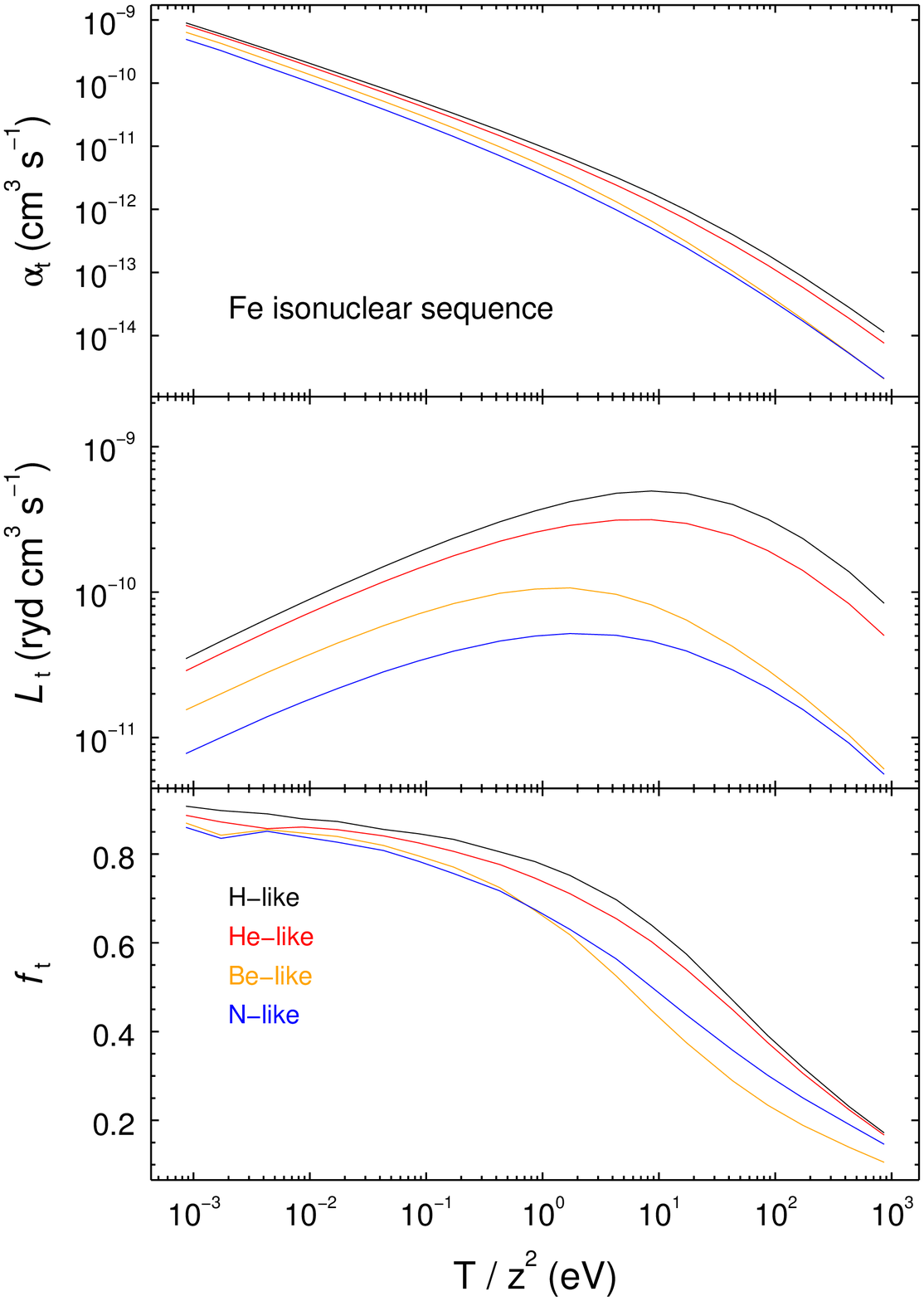}
\caption{Top panel is total RR rates $\alpha_{\rm t}$ of the Fe isonuclear sequence, 
including H- (black), He- (red), Be- (orange) and N-like (blue); 
Middle panel is the RR electron energy-loss rates $L_{\rm t}$; 
And bottom panel is the weighted electron energy-loss factors $f_{\rm t}$. 
The temperature of the plasma is down-scaled by $z^2$, as in Figure~\ref{fig:rr_welf_02ies}.
}
\label{fig:rr_welf_ins26}
\end{figure}

The bottom panels of Figure~\ref{fig:rr_welf_02ies} 
and \ref{fig:rr_welf_ins26} illustrate the weighted electron energy-loss factors 
for He-like isoelectronic sequences (He, Si and Fe) 
and Fe isonuclear sequence (H-, He-, Be- and N-like), respectively.
The deviation from (slightly below) unity at the lower temperature end is simply 
due to the fact that the weighted electron energy-loss factors of 
the very high-$n$ shells are no longer close to unity 
(Figure~\ref{fig:rr_welf_0426}, middle panel).
The deviation from (slightly above) zero at the high temperature end is 
because the ionization potentials of the first few low-$n$ shells can 
still be comparable to $kT$, while sum of these $n$-resolved RR rates 
are more or less a few tens of percent of the total RR rates. 

Due to the non-hydrogenic screening of the wave function 
for low-$nl$ states in low-charge many-electron ions,
the characteristic high-temperature bump is present
in not only the RR rates \citep[see Figure~4 in][for an example]{bad06}
but also in the electron energy-loss rates.
The feature is even enhanced in the weighted electron energy-loss factor. 

\subsection{Parameterization}
\label{sct:par}
We parameterize the ion/atom-resolved radiative recombination electron energy-loss factors 
using the same fitting strategy described in \citet{mao16}, with the model function of  
\begin{equation}
\small
f_{\rm t}(T) = a_{0}~T^{-b_{0} - c_{0}\log{T}}~\left(\frac{1 + a_2 T^{-b2}}{1 + a_1 T^{-b1}}\right)~,
\label{eq:fit_m16}
\end{equation}
where the electron temperature $T$ is in units of eV,
$a_{0}~{\rm and}~b_{0}$ are primary fitting parameters,
$c_0,~a_{1,~2}$ and $b_{1,~2}$ are additional fitting parameters.
The additional parameters are frozen to zero if they are not used.
Furthermore, we constrain $b_{0-2}$ to be within -10.0 to 10.0
and $c_0$ between 0.0 and 1.0.
The initial values of the two primary fitting parameters
$a_{0}~{\rm and}~b_{0}$ are set to unity
together with the four additional fitting parameters
$a_{1,~2}~{\rm and}~b_{1,~2}$ if they are thawn.
Conversely, the initial value of $c_0$, if it is thawn,
is set to either side of its boundary,
i.e. $c_0 = 0.0$ or $c_0 = 1.0$ (both fits are performed).

In order to estimate the goodness of fit,
the fits are performed with a set of artificial relative errors $(r)$.
We started with $r=0.625\%$, 
following with increasing the artificial relative error by a factor of two,
up to and including $2.5\%$.
The chi-squared statistics adopted here are
\begin{equation}
\small
\chi^2 = \sum_{i = 1}^{N} \left(\frac{n_i - m_i}{r~n_i}\right)^2~, 
\label{eq:chi2}
\end{equation}
where $n_i$ is the $i$th numerical calculation result and $m_i$ is the $i$th
model prediction (Equation~\ref{eq:fit_m16}). 

For the model selection, we first fit the data with the simplest model 
(i.e. all the five additional parameters are frozen to zero), 
following with fits with free additional parameters step by step. 
Thawing one additional parameter decreases the degrees of freedom by one, 
thus, only if the obtained statistics ($\chi^2$) of the more complicated model
improves by at least 2.71, 4.61, 6.26, 7.79 and 9.24 
for one to five additional free parameter(s), respectively, 
the more complicated model is favored (at a 90\% nominal confidence level). 

\begin{figure}
\centering
\includegraphics[width=\hsize]{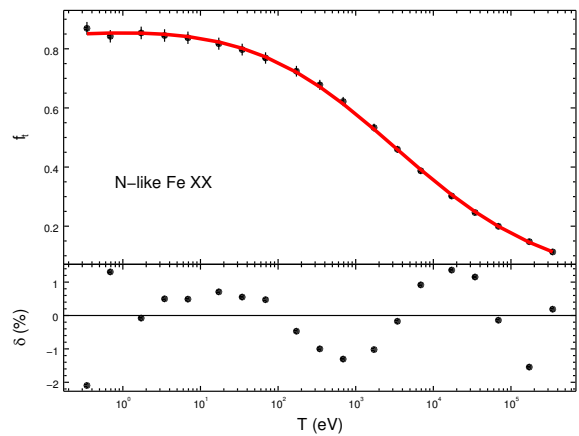}
\caption{The RR weighted electron energy-loss factor for N-like iron (\ion{Fe}{xx}).
The black dots in both panels (associated with artificial error bars of 2.5\% in the upper one) 
are the calculated weighted electron energy-loss factor. 
The red solid line is the best-fit.
The lower panel shows the deviation (in percent) 
between the best-fit and the original calculation.
}
\label{fig:rr_welf_0726}
\end{figure}

Parameterizations of the ion/atom-resolved RR weighted electron energy-loss factors 
for individual ions/atoms in H-like to Ne-like isoelectronic sequences were performed. 
A typical fit for non-hydrogenic systems is shown in 
Figure~\ref{fig:rr_welf_0726} for N-like iron (\ion{Fe}{xx}).
The fitting parameters can be found in Table~\ref{tbl:fit_case}.
Again, the weighted energy-loss factor per ion/atom is 
close to unity at low temperature end and drops towards zero rapidly 
at the high temperature end. 

\begin{figure}
\centering
\includegraphics[width=\hsize]{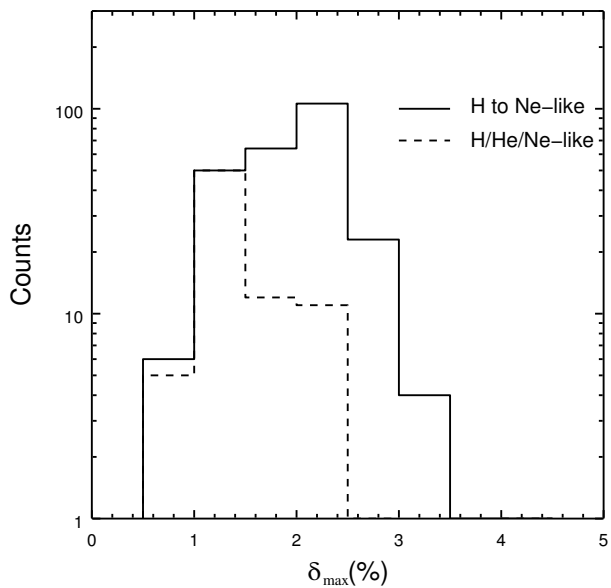}
\caption{The histogram of maximum deviation in percent ($\delta_{\rm max}$) 
for all the ions considered here, which reflects the overall goodness of
our parameterization. The dashed-histogram is the statistics of the more important 
H-like, He-like and Ne-like isoelectronic sequences, 
while the solid one is the statistics of all the isoelectronic sequences.
}
\label{fig:welf_ioa_stat}
\end{figure}

In Figure~\ref{fig:welf_ioa_stat} we show the histogram of 
maximum deviation $\delta_{\rm max}$ (in percent)
between the fitted model and the original calculation for all the ions considered here. 
In short, our fitting accuracy is within 4\%, and even accurate ($\lesssim 2.5\%$)
for the more important H-like, He-like and Ne-like isoelectronic sequences.

\begin{figure}
\centering
\includegraphics[width=\hsize]{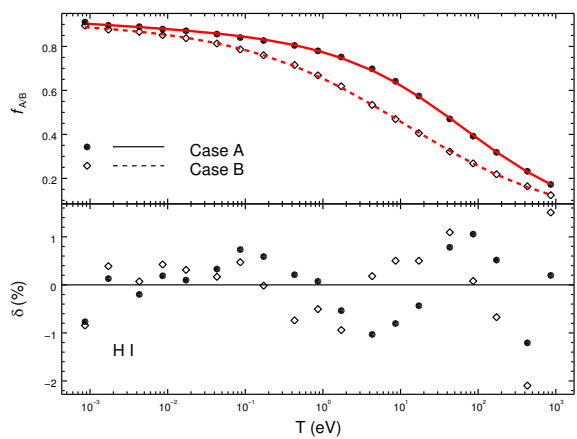}
\caption{The Case A (solid line, filled circles) and Case B (dashed line, empty diamonds) 
RR weighted electron energy-loss factor ($f_{\rm A/B}$) for \ion{H}{i}.
The black dots in both panels (associated with artificial error bars in the upper one) 
are the calculated weighted electron energy-loss factor. 
The red solid line is the best-fit.
The lower panel shows the deviation (in percent) 
between the best-fit and the original calculation.
}
\label{fig:rr_welf_0101case}
\end{figure}

\begin{figure}
\centering
\includegraphics[width=\hsize]{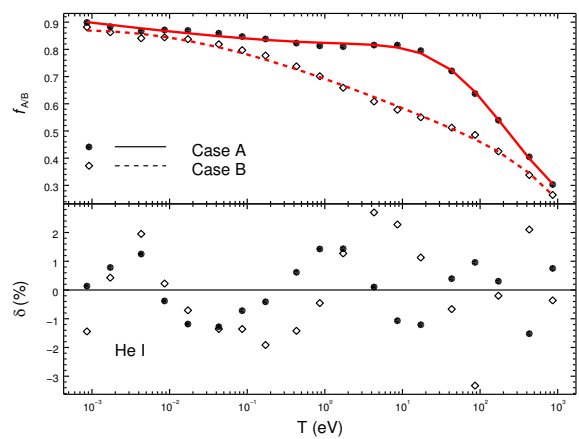}
\caption{Similar to Figure~\ref{fig:rr_welf_0101case} but for  \ion{He}{i}.
}
\label{fig:rr_welf_0202case}
\end{figure}

In addition, we also fit specifically the Case A
($f_{\rm A} = \beta_{\rm t}/\alpha_{\rm t}$) 
and Case B \citep[][$f_{\rm B} = \beta_{n\ge2}/\alpha_{n\ge2}$]{bak38}
RR weighted electron energy-loss factors of \ion{H}{i} (Figure~\ref{fig:rr_welf_0101case}) 
and \ion{He}{i} (Figure~\ref{fig:rr_welf_0202case}).
Typical unparameterized factors ($f_{\rm A}$ and $f_{\rm B}$)
and fitting parameters can be found 
in Table~\ref{tbl:unfit_case} and \ref{tbl:fit_case}, respectively.


\begin{table}
\caption{Unparameterize of RR weighted electron energy-loss factors 
for \ion{H}{i}, \ion{He}{i} and \ion{Fe}{xx}.
For the former two, both Case A and Case B results are treated seperately.}
\label{tbl:unfit_case}
\centering
\begin{tabular}{cccccccccccccccccccccc}
\hline\hline
 $T/z^2$ & \ion{H}{i}  & \ion{H}{i} &  \ion{He}{i} & \ion{He}{i} & \ion{Fe}{xx}  \\
 K & Case A & Case B & Case A & Case B & Case A \\
\hline\hline
$10^1$ & 0.911 & 0.895 & 0.899 & 0.882 & 0.869 \\
$10^2$ & 0.879 & 0.851 & 0.871 & 0.844 & 0.845 \\
$10^3$ & 0.841 & 0.786 & 0.847 & 0.797 & 0.797 \\
$10^4$ & 0.780 & 0.668 & 0.813 & 0.701 & 0.678 \\
$10^5$ & 0.642 & 0.470 & 0.816 & 0.578 & 0.460 \\
$10^6$ & 0.392 & 0.268 & 0.637 & 0.486 & 0.246 \\
$10^7$ & 0.172 & 0.123 & 0.303 & 0.265 & 0.113 \\
 \hline
\end{tabular}
\tablefoot{Machine readable unparameterized Case A factors
for all the ions considered here are available on CDS.
}
\end{table}

\begin{table*}
\caption{Fitting parameters of RR weighted electron energy-loss factors 
for \ion{H}{i}, \ion{He}{i} and \ion{Fe}{xx}.
For the former two, both Case A and Case B results are included.}
\label{tbl:fit_case}
\centering
\begin{tabular}{cccccccccccccccccccccc}
\hline\hline
$s$ & $Z$ & Case & $a_0$ & $b_0$ & $c_0$ & $a_1$ & $b_1$ & $a_2$ & $b_2$ & $\delta_{\rm max}$  \\
\hline\hline
 1 &  1 & A & 8.655E+00 & 5.432E-01 & 0.000E+00 & 1.018E+01 & 5.342E-01 & 0.000E+00 & 0.000E+00 & 1.2\% \\ 
 1 &  1 & B & 2.560E+00 & 4.230E-01 & 0.000E+00 & 2.914E+00 & 4.191E-01 & 0.000E+00 & 0.000E+00 & 2.1\% \\ 
 2 &  2 & A & 2.354E+00 & 3.367E-01 & 0.000E+00 & 6.280E+01 & 8.875E-01 & 2.133E+01 & 5.675E-01 & 1.5\% \\ 
 2 &  2 & B & 1.011E+04 & 1.348E+00 & 4.330E-03 & 1.462E+04 & 1.285E+00 & 0.000E+00 & 0.000E+00 & 3.5\% \\ 
 7 & 26 & A & 2.466E+01 & 4.135E-01 & 0.000E+00 & 2.788E+01 & 4.286E-01 & 0.000E+00 & 0.000E+00 & 2.1\% \\ 
 \hline
\end{tabular}
\tablefoot{$s$ is the isoelectronic sequence number of the recombined ion, 
$Z$ is the atomic number of the ion, 
$a_{0-2}$, $b_{0-2}$ and $c_0$ are the fitting parameters
and $\delta_{\rm max}$ is the maximum deviation (in percent) between the ``best-fit"
and original calculation.
Case A and Case B refers to $\beta_{\rm t}/\alpha_{\rm t}$
and $\beta_{n\ge2}/\alpha_{n\ge2}$
RR weighted electron energy-loss factors, respectively.
Machine readable fitting parameters and maximum deviation (in percent) 
for the total weighted electron energy loss factors 
for all the ions considered here are available on CDS.
}
\end{table*}

\section{Discussions}
\label{sct:dis}
\subsection{Comparison with previous results for \ion{H}{i} and \ion{He}{i}}
\label{sct:cf_0101_0202}
\begin{figure}
\centering
\includegraphics[width=\hsize]{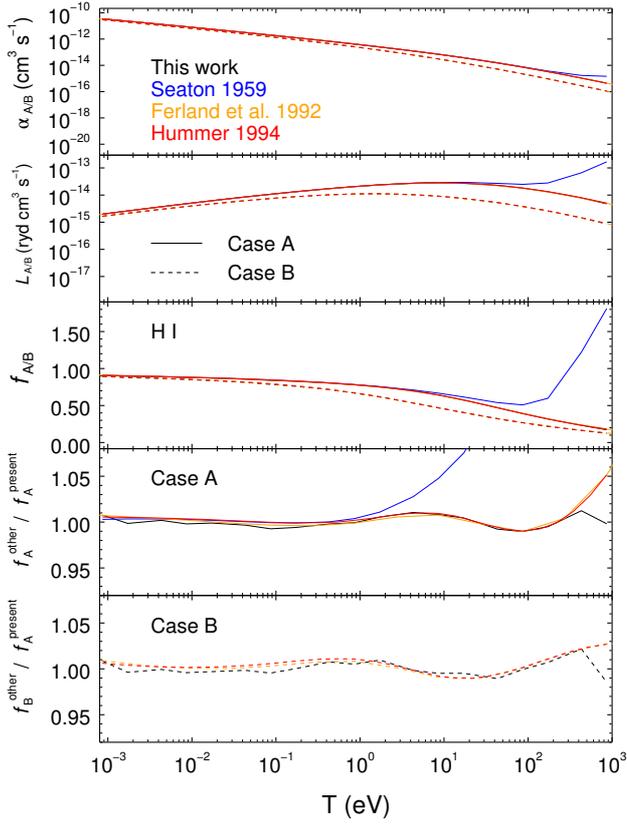}
\caption{The comparison of the RR data for \ion{H}{i} 
among results from this work (black), \citet[][blue]{sea59}, 
\citet[][orange]{fer92} and \citet[][red]{hum94}. 
Both results of case A (solid lines) 
and case B (dashed lines) are shown.
The total RR rates ($\alpha_{\rm A/B}^{\rm RR}$) and 
electron energy-loss rates ($L_{\rm A/B}^{\rm RR}$) 
are shown in the top two panels. 
The RR weighted electron energy-loss factors ($f_{\rm A/B}$) are 
shown in the middle panel. 
The ratios of $f_{\rm A/B}$ from this work and previous works 
with respect to the fitting results (Equation~\ref{eq:fit_m16} 
and Table~\ref{tbl:fit_case}) of this work, 
i.e. ${f_{\rm A/B}^{\rm other}}/{f_{\rm A/B}^{\rm present}}$,
are shown in the bottom two panels.}
\label{fig:0101_cf_ref}
\end{figure}

Figure~\ref{fig:0101_cf_ref} shows a comparison of RR rates ($\alpha_{t}^{\rm RR}$), 
electron energy-loss rates ($L_{t}^{\rm RR}$), 
weighted electron energy-loss factors ($f_{t}^{\rm RR}$) from this work, 
\citet[][blue]{sea59}, \citet[][orange]{fer92} and \citet[][red]{hum94}. 
Since both \citet{fer92} and \citet{hum94} use the same PICSs \citep{sto91}, 
the two results are highly consistent as expected. 
The Case A and Case B results of this work are also consistent within 1\% 
at the low temperature end, and increase to $\sim$5\% (underestimation).
For the high temperature end ($T\gtrsim 0.1$~keV), 
since the ion fraction of \ion{H}{i} is rather low (almost completely ionized), 
the present calculation is still acceptable. 
A similar issue towards to the high temperature end 
is also found in the Case A results of \citet{sea59}, 
with a relatively significant overestimation ($\gtrsim5\%$) from the other three calculations.

\begin{figure}
\centering
\includegraphics[width=\hsize]{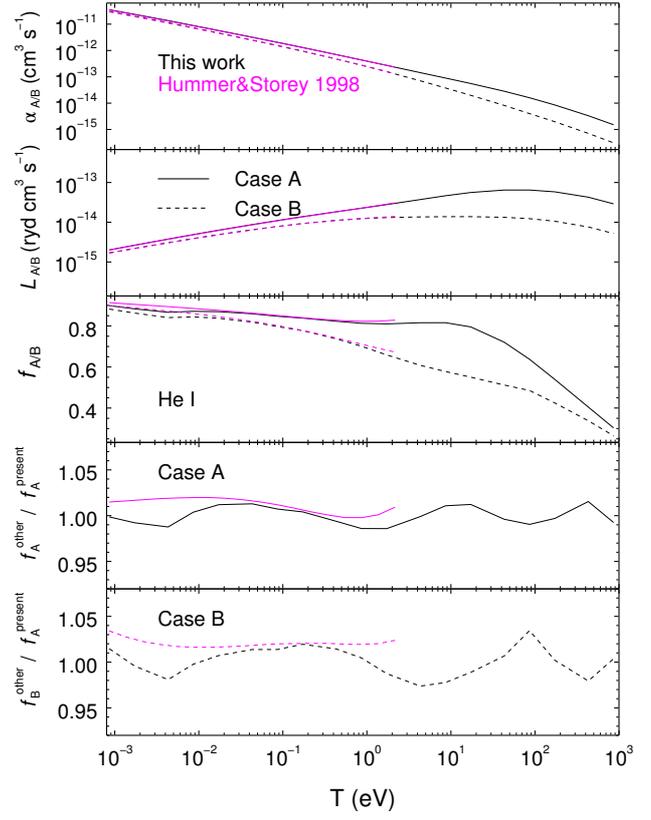}
\caption{Similar to Figure~\ref{fig:0202_cf_ref} but for \ion{He}{i} 
between this work (black) and \citet[magenta][]{hum98}. 
The latter one only provides data with $T \le 10^{4.4}~{\rm K}$.
}
\label{fig:0202_cf_ref}
\end{figure}

Likewise, the comparison for \ion{He}{i} between this work and 
\citet{hum98} is presented in Figure~\ref{fig:0202_cf_ref}.
The Case A and Case B results from both calculations agree well (within $2\%$), 
at the low temperature end ($T\lesssim 2.0$~eV). 
At higher temperatures with $T\gtrsim 2$~eV,
the RR rate and electron energy-loss rate for \ion{He}{i}  
are not available in \citet{hum98}.

\subsection{Scaling with $z^2$}
\label{sct:sca_z2}
Previous studies of hydrogenic systems, \citet{sea59, fer92, hum98}, 
all use $z^2$ scaling for $\alpha_{\rm t}^{\rm RR}$.
That is to say, $\alpha_{\rm t}^{X}= z^2~\alpha_{\rm t}^{\rm H}$,
where $z$ is the ionic charge of the recombined ion $X$.
The same $z^2$ scaling also applies for  $\beta_{\rm t}^{\rm RR}$ (or $L_{\rm t}^{\rm RR}$). 
\citet{lam01} also pointed out that the shell-resolved ratio of $f_{n}^{RR}$
(=$\beta_{n}^{\rm RR}/\alpha_{n}^{\rm RR}$) can also be scaled with $z^2 / n^2$,
i.e. $f_{n}^{X}= \frac{z^2}{n^2}~f_{n}^{\rm H}$
with $n$ refers to the principle quantum number.

\begin{figure}
\centering
\includegraphics[width=\hsize]{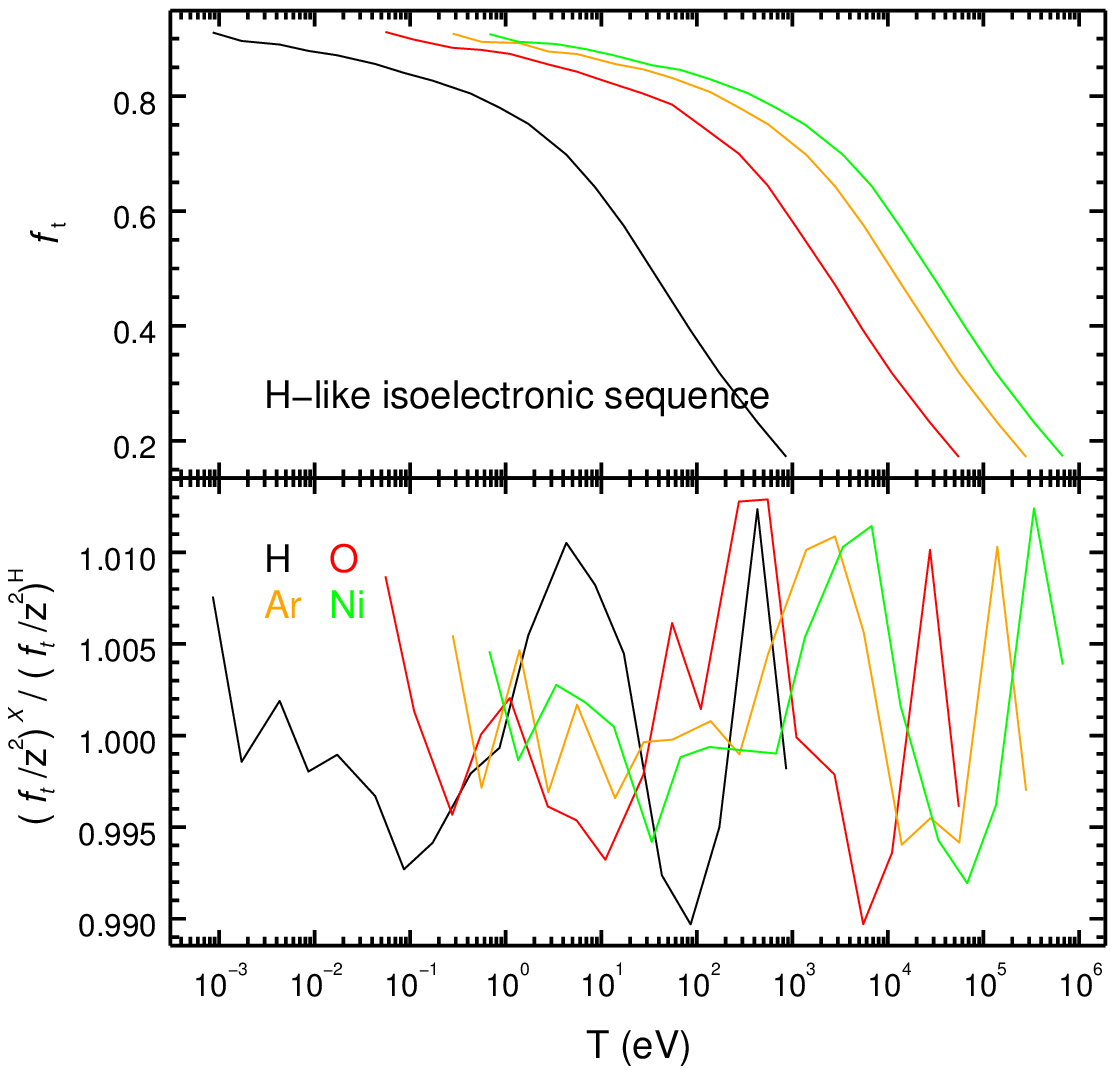}
\caption{The $z^2$ scaling for the H-like isoelectronic sequence (Case A), 
including \ion{H}{i} (black), \ion{O}{viii} (red), \ion{Ar}{xviii} (orange) and \ion{Ni}{xxviii} (green). 
The top panel shows the ratios of $f_t/z^2$ as a function of electron temperature ($T$).
The bottom panel is the ratio of $(f_{\rm t}/z^2)^{X}$ for ion $X$
with respect to the ratio of $(f_{\rm t}/z^2)^{\rm H}$ for H.
}
\label{fig:welf_01ies_sca}
\end{figure}

\begin{figure}
\centering
\includegraphics[width=\hsize]{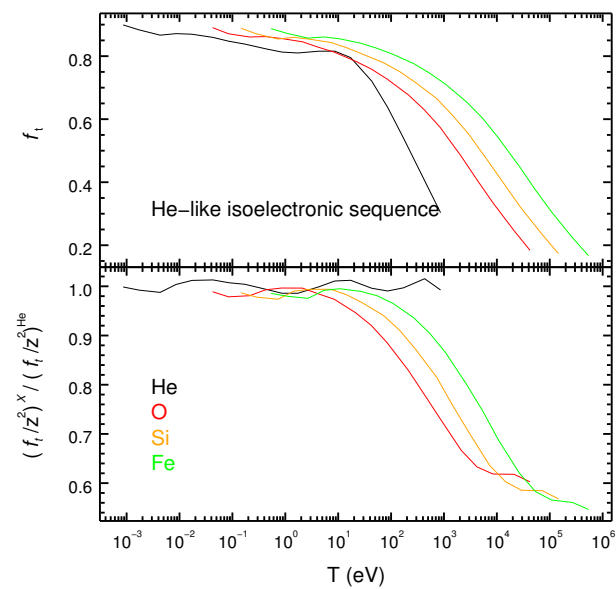}
\caption{Similar to Figure~\ref{fig:welf_01ies_sca}
but for the $z^2$ scaling for the He-like isoelectronic sequences.
}
\label{fig:welf_02ies_sca}
\end{figure}

\begin{figure}
\centering
\includegraphics[width=\hsize]{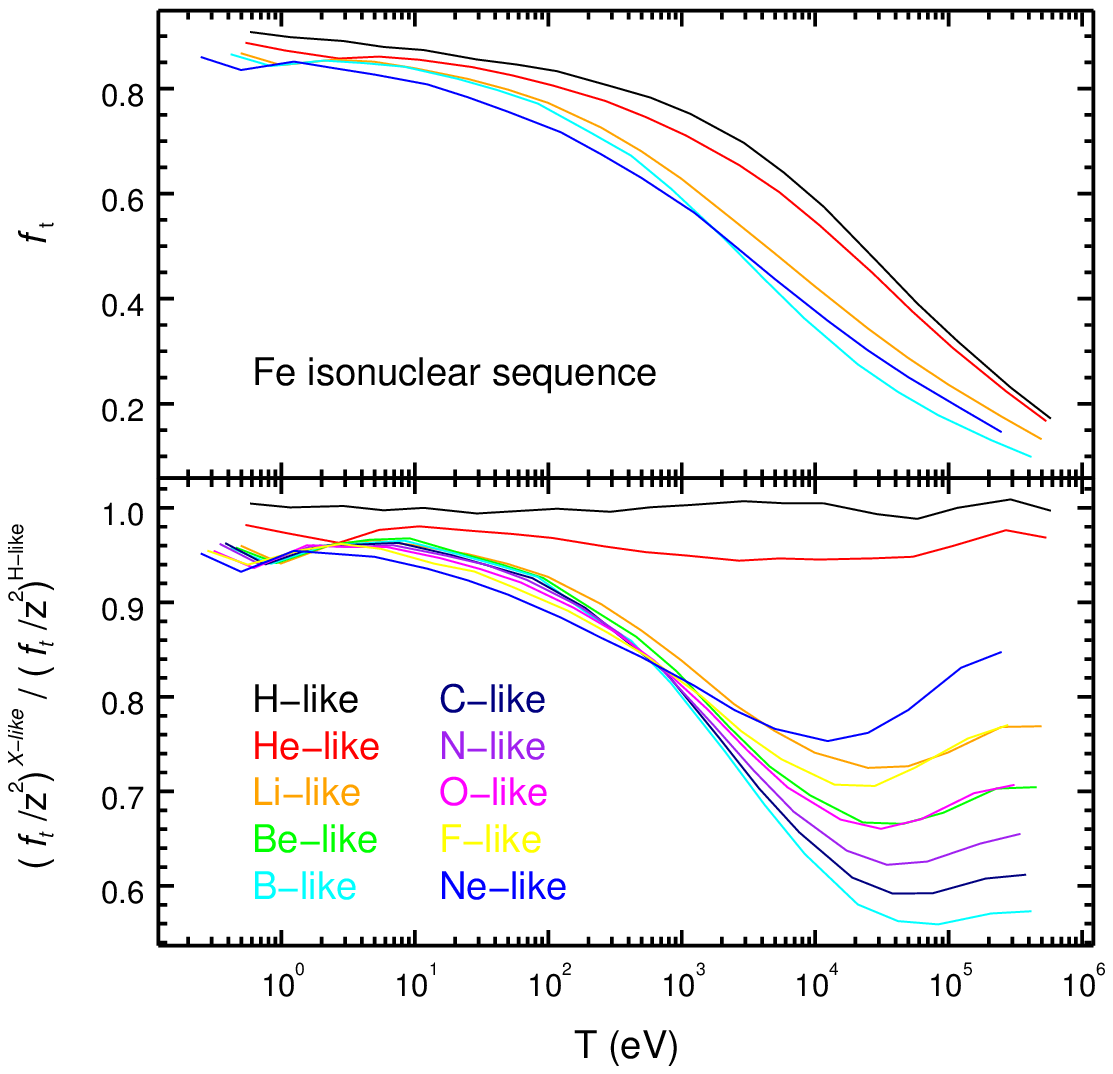}
\caption{The $z^2$ scaling for the Fe isonuclear sequence. 
The top panel shows the ratios of $f_{\rm t}/z^2$ 
as a function of electron temperature ($T$). 
The bottom panel is the ratio of $(f_{\rm t}/z^2)^{X-{\rm like}}$ for $X$-like Fe  
with respect to the ratio of $(f_{\rm t}/z^2)^{\rm H-like}$ for H-like \ion{Fe}{xxvi}.}
\label{fig:welf_ins26_sca}
\end{figure}

In the following, we merely focus on the scaling for the ion/atom-resolved data set.
We show in the top panel of Figure~\ref{fig:welf_01ies_sca} 
the ratios of $f_t/z^2$ for H-like ions. Apparently, 
from the bottom panel of Figure~\ref{fig:welf_01ies_sca}, 
the $z^2$ scaling for the H-like isoelectronic sequence is accurate within 2\%.
For the rest of the isoelectronic sequences, for instance, 
the He-like isoelectronic sequence shown in Figure~\ref{fig:welf_02ies_sca},
the $z^2$ scaling applies at the low temperature end, 
whereas, the accuracies are poorer toward the high temperature end.
We also show the $z^2$ scaling for the Fe isonuclear sequence 
in Figure~\ref{fig:welf_ins26_sca}. 

\subsection{Radiative recombination continua}
\label{sct:rrc}
\begin{figure}
\centering
\includegraphics[width=\hsize]{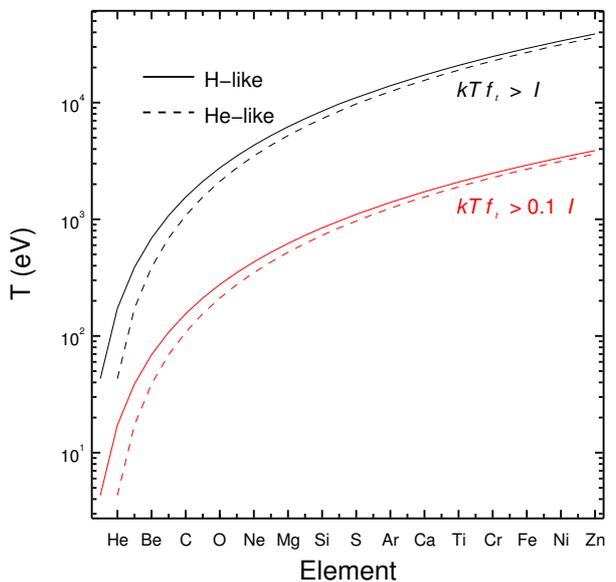}
\caption{The threshold temperature above which the electron energy-loss via RR 
cannot be neglected, compared to the ion energy-loss, for H-like (solid lines) 
and He-like ions (dashed lines). 
}
\label{fig:rr_welf_rrc}
\end{figure}

We restrict the discussion above
for the RR energy-loss of the electrons in the plasma only.
The ion energy-loss of the ions due to RR can be estimated as
$P^{\rm RR} \sim I_i~\alpha_i$,
where $I_i$ is the ionization potential of
the level/term the free electron is captured into,
and $\alpha_i$ is the corresponding RR rate coefficient.
Whether to include the ionization potential energies as part of
the total internal energy of the plasma is not critical,
as long as the entire computation of the net energy gain/loss is self-consistent
\citep[see a discussion in ][]{gna12}.
On the other hand, when interpreting the emergent spectrum due to RR,  
such as the radiative recombination continua (RRC) for a low-density plasma,
the ion energy-loss of the ion is essentially required. 
The RRC emissivity \citep{tuc66} can be obtained via
\begin{eqnarray}
\small
\frac{dE^{\rm RRC}}{dt~dV} &=& \int_{0}^{\infty} n_{\rm e}~n_{\rm i}~\left(I + \frac{1}{2} m v^2\right)~v~\sigma(v)~f(v,~kT) dv \nonumber \\ 
&=& n_{\rm e}~n_{\rm i}I~\left(1 + f_{\rm t}~kT/I\right)~\alpha_{\rm t}~,
\label{eq:rrc}
\end{eqnarray}
where $n_{\rm e}$ and $n_{\rm i}$ are the electron and (recombining) ion number density,
respectively. Generally speaking, the ion energy-loss of the ion dominates 
the electron energy-loss of the electrons,
since $f_{\rm t}$ is of the order of unity while $kT \lesssim I$ holds
for those X-ray photoionizing plasmas in XRBs \citep{lie96},
AGN \citep{kin02} and recombining plasmas in SNRs \citep{oza09}. 
Figure~\ref{fig:rr_welf_rrc} shows the threshold temperature above which 
the electron energy-loss via RR cannot be neglected 
compared to the ion energy-loss.
For hot plasmas with $kT \gtrsim 2$~keV, 
the electron energy-loss is comparable to the ion energy-loss for $Z>5$. 
It is necessary to emphasize that we refer to 
the electron temperature $T$ of the plasma here,
which is not necessarily identical to the ion temperature of the plasma,
in particular, in the nonequilibrium ionization scenario. 

\subsection{Total radiative recombination rate}
\label{sct:rr_tot}
Various calculations of (total or shell/term/level-resolved) RR data are available from the literature.
Historically, different approaches have been used for calculating the total RR rates,
including the Dirac-Hartree-Slater method \citep{ver93}
and the distorted-wave approximation \citep{gu03, bad06}.
Additionally, Nahar and coworkers \citep[e.g.][]{nah99} obtained 
the total (unified DR + RR) recombination rate for various ions 
with their \textbf{\textit{R}}-matrix calculations.
Different approaches can lead to different total RR rates \citep[see a discussion in][]{bad06}, 
as well as the Individual term/level-resolved RR rate coefficients, 
even among the most advanced \textbf{\textit{R}}-matrix calculations.
Nevertheless, the bulk of the total RR rates for various ions 
agrees well among each other. 
As for the detailed RR rate coefficients, 
consequently, the detailed RR electron energy-loss rate, 
as long as the difference among different methods are within a few percent 
and given the fact that each individual RR is $\lesssim 10\%$ of 
the total RR rate for a certain ion/atom, 
the final difference in the total weighted electron energy-loss factors
$f_{\rm t}$ are still within 1\%.   
In other words, although we used the re-calculated total RR rate 
(Section \ref{sct:rc}) to derive the weighted electron energy-loss factors, 
we assume these factors can still be applied to other total RR rates.

\begin{acknowledgements}
J.M. acknowledges discussions and support from 
M. Mehdipour, A. Raassen, L. Gu and M. O'Mullane.
We thank the referee, G. Ferland, 
for the valuable comments on the manuscript
SRON is supported financially by NWO, 
the Netherlands Organization for Scientific Research.
\end{acknowledgements}


\end{document}